\documentclass[11pt]{article}

\usepackage[utf8]{inputenc}
\usepackage[T1]{fontenc}
\usepackage{lmodern}
\usepackage{amsmath,amssymb}
\usepackage{booktabs}
\usepackage{graphicx}
\usepackage{geometry}
\usepackage{hyperref}
\usepackage{xcolor}
\usepackage{enumitem}
\usepackage[ruled,vlined,linesnumbered]{algorithm2e}
\usepackage{microtype}
\usepackage[font=small,labelfont=bf,skip=4pt]{caption}
\usepackage{placeins}

\hypersetup{colorlinks=true, linkcolor=blue!50!black, citecolor=blue!50!black,
            urlcolor=blue!50!black}

\newcommand{\pov}{\textsc{Shadow-PPOV}}

\title{\textbf{Model-Free Passive Execution via Order-Level Shadowing}}

\author{%
  Vincent Maciejewski\\
  \small M2 Technologies\\
  \small \texttt{mayeski@gmail.com}
}

\date{September 2026}

\begin{document}
\maketitle

\begin{abstract}
Automated execution algorithms are organized into schedule-based and
liquidity-seeking families \cite{johnson2010,bayesiantca2019}. This paper
concerns the first, whose members --- Time-Weighted Average Price (TWAP),
Volume-Weighted Average Price (VWAP), Percentage of Volume (POV) and
Implementation Shortfall --- are all \emph{model-based}: each derives its
decisions from an explicit model, forecast, schedule or control rule. We introduce \pov{}, a passive POV
whose \emph{order-placement rate} is set from observed order flow rather than
from traded volume. Placing a passive order to fill efficiently conventionally
involves an order-book model and a fill prediction.
\pov{} replaces that prediction with \emph{tracking}: on observing a
third-party add, it may transmit its own limit order at the same price
\emph{on the same venue}, recording a single association between the observed
order's exchange identifier and its own. Cancellation is then
\emph{identifier-driven} --- the shadow is withdrawn when the order it follows
ends, at once on a cancel and after a brief grace window on a trade. The placement decision is thus \emph{model-free}: price
and venue are read off the observed order.

Model-free is not information-free. \pov{} reads every order-book message and
places only where a participant has just committed capital, while
\emph{computing} nothing from what it reads. Information is inherited from the
flow rather than derived from a model.

We evaluate \pov{} on a full calendar year of replayed Chicago Mercantile
Exchange (CME) ES futures in a deterministic market-replay simulator, reporting
its slippage and latency sensitivity and comparing it against the aggressive
equivalent POV. We propose it as a model-free benchmark for passive-order
placement, against which a predictive placement model can be scored. The
algorithm and the order-book simulator are implemented in the open-source
\textsc{kaspar-hft} project.\footnote{\url{https://github.com/vincent212/kaspar-hft}}
\end{abstract}

\vspace{0.5em}

\section{Introduction}

A large order cannot be executed in a single trade without paying an impact
cost proportional to how much liquidity it consumes at once. The standard
response is to slice the parent order into child orders spread over time. The
question every execution algorithm must answer is \emph{how to slice}: 
how much to trade, when, and at what price.

TWAP commits to
a clock: it slices evenly in time, blind to where liquidity actually is. 
But it must decide how much of each slice to work passively rather 
than cross the spread, and that tactical choice is shared across 
schedule-based algorithms rather than tied to any one of them \cite{scheduletactics2014,limitmarkettactics2014}.
VWAP
commits to a \emph{forecast of the intraday volume curve} and weights its
slices to match it; the forecast error is the tracking error.
Percentage-of-volume (POV) has a participation ratio \emph{reactive} to
realized volume ---
it does not forecast a volume curve the way VWAP does
\cite{gueantpov2012,kissell}.
But POV cannot participate \emph{passively} by reaction alone,
because volume that has already printed can no longer be joined at a resting
price; to stay passive it must lean slightly ahead of near-term volume (or
cross the spread to catch up \cite{cartea2015,contkukanov2015}), and its completion time depends on volume it
cannot yet see \cite{vaeshauser2022}. So POV trades a volume-\emph{curve} forecast for a
short-horizon anticipation plus completion uncertainty, and it tracks an
\emph{aggregate} volume statistic rather than the individual resting orders
that make up the book.

Implementation Shortfall takes a different benchmark: the arrival price, the
market price at the moment the parent order is released. It measures execution
cost as the gap between that decision price and the realized average fill price,
including the opportunity cost of any unfilled quantity. Minimizing the gap
trades market impact, which favors slow trading, against timing risk from price
drift and volatility, which favors fast trading. The resulting schedule is
front-loaded and tuned by a risk-aversion parameter: higher risk aversion trades
faster to reduce exposure to adverse drift, at the cost of more impact. Like the volume-based algorithms, 
IS trades to a schedule computed in advance from a volume and
impact forecast. Perold \cite{perold1988} introduces the IS formulation, and
Almgren--Chriss \cite{almgren2000} give the arrival-price optimal-execution
formulation IS approximates.

This work concerns \emph{passive} execution specifically, so the way POV places
passive orders is central. A conventional POV algorithm is a hybrid: it rests
passive limit orders when the market allows, but to hold its participation rate
it also sends marketable orders, and it crosses the spread to catch up when it
falls behind or when a slice deadline arrives \cite{cartea2015,contkukanov2015}.
The passive-versus-aggressive choice is not special to POV: schedule-based
implementations commonly face the additional tactical question of how much of
each slice to work passively rather than
cross, and this tactical layer is shared across schedules rather than tied to any one algorithm \cite{scheduletactics2014,limitmarkettactics2014}. A schedule
that crossed on every slice would pay the spread needlessly, so even TWAP
typically posts each slice passively and crosses only the unfilled residual near
the slice deadline \cite{cesari2012}.

Wherever a slice is worked passively, the same placement problem arises: how far
into the book to quote, trading fill probability against market impact and
adverse selection \cite{glostenmilgrom1985,gueant2012,passiveimpact2026,negdrift2024}. A passive order can
simply be posted at the touch, which needs no model; but posting blindly at the
touch surrenders control of that trade-off and is prone to adverse selection ---
the order fills exactly when the level is thinning ahead of an adverse move.
The conventional solutions instead
build an order-book model and predict where a passive order will fill:
optimal placement of limit versus market orders \cite{cartea2015}, optimal
liquidation with limit orders under a fill-intensity model \cite{gueant2012},
and convex optimization of order placement across the book \cite{contkukanov2015}.
\pov{} removes the need for an explicit order-book model and fill prediction from
the placement decision: it resolves the placement
problem by \emph{following}, resting each passive order at the price of an order
that just arrived at the book. This replaces order-book modeling and fill prediction with
tracking of where liquidity actually rests. \pov{} governs the placement of
passive orders only; how much of a parent order to work passively versus
aggressively is a separate scheduling decision. Production implementations may
mix passive and aggressive orders, and this paper makes no claim that aggressive
orders are inferior. We compare passive Shadow-PPOV to aggressive POV
in Section~\ref{sec:eval:results}.

\paragraph{Model-free is not information-free.} The two are easily conflated and
the difference decides what this method is. \pov{} is saturated in market
information: it consumes every add, cancel and trade message on its side of the
book, and it acts only where another participant has just committed capital at a
price. A resting limit order is somebody's conclusion, arrived at by whatever
means, that this price is worth standing at. Shadowing it inherits that
conclusion. The whole of \pov{}'s own contribution is to copy that price and
watch that order's identifier. Everything a placement model would normally
supply --- a picture of the book, a fill probability, a value for queue
position, a view on where the price is going --- is simply absent from the
decision. It free-rides on the modelling other participants have already done
rather than performing any of its own, which is why it can be specified in a
page (Algorithm~\ref{alg:core}) and why its cancellation decision needs no state
beyond an identifier.

This also bounds what the method can be expected to do. Inheriting a
\emph{placement} judgement is not the same as holding a \emph{directional} view:
knowing where a fill is likely says nothing about where the price is going next.
Section~\ref{sec:disc} argues that a passive method without a directional
forecast should execute slightly worse than the mid-price whatever else it
knows, and Section~\ref{sec:eval} measures exactly that. The latency results of
Section~\ref{sec:latency} then show the inherited information is real and
perishable: its value decays as the delay between observing an order and acting
on it grows.

Its model-free character sets \pov{} apart from
standard execution algorithms, and lets it serve as a
baseline against which predictive placement models can be measured
(Section~\ref{sec:disc}).

Beyond the question of how much to post and when, conventional execution
stacks carry two further sources of overhead that this work targets directly:

\begin{itemize}[leftmargin=1.4em]
  \item \textbf{Routing computation.} Smart order routing (SOR) across venues
  is normally a calculation: aggregate per-venue book depth, compute venue
  attractiveness scores, model per-venue latency, price maker/taker fees, and
  optimize a routing objective.
  \item \textbf{State-dependent cancellation.} Order-management systems
  typically decide whether to pull a resting order by continuously evaluating
  market state --- spread, queue position, fill probability, price
  competitiveness --- which requires maintaining substantial per-order state
  and adds decision latency.
\end{itemize}

The premise of this paper is that all three --- the forward placement of passive
orders, the routing computation, and the state-dependent cancellation --- are
avoidable for a passive execution mandate, and that avoiding them yields a
simple, computationally light method. The algorithm
does not need to predict anything; it only has to follow the market. \pov{}
replaces the forward commitment with an observation: the passive orders other
participants have \emph{already committed to the book}.
In the price-time-priority (FIFO) model used here, a limit order placed at the
same price and venue as a resting participant order joins behind that order in
the queue; a sweep must therefore consume the participant order and the depth
ahead of the shadow before reaching the shadow. The
method replaces the routing calculation with a pass-through of the observed
order's originating venue, and replaces state-dependent cancellation with an
identifier rule: the shadow is withdrawn when the order it follows ends.

\paragraph{A note on the word ``shadow''.} Here it means \emph{attached to}. A
shadow order is an ordinary displayed limit order, bound by identifier to a
particular third-party order and following that order's price, venue and
lifecycle --- it shadows the way a person shadows another, by going where they
go. Some venues also use ``shadow'' as a label for hidden or reserve order
types, which is a different word doing a different job; we flag it only so the
reader reads ours in the right sense. Every order \pov{} sends is displayed.

\paragraph{Contributions.}
\begin{enumerate}[leftmargin=1.6em]
  \item We propose \pov{} as a simple, \emph{model-free} benchmark for
  passive-order placement, built on \emph{participant shadowing}: on observing a
  third-party add, \pov{} places a passive order at that add's price and venue,
  attaches it to the arriving order by exchange identifier, and ties its
  lifecycle to that order. Placement is model-free in the sense used
  throughout: price and venue are read off the observed order, giving predictive
  placement models a reference to beat (Sections~\ref{sec:algo},
  \ref{sec:disc}).
  \item We define \emph{identifier-driven cancel mirroring}: cancellation
  triggered by a delete/trade for the followed order's identifier rather than
  by market-state evaluation, retaining a single identifier association per
  followed order --- the exchange order identifier paired with the shadow's
  --- so that no book state is available to the cancellation decision even in
  principle (Sections~\ref{sec:map}--\ref{sec:cancel}).
  \item We \emph{propose} \,--- and argue analytically, without evaluating
  \,--- \emph{flow-inherited venue selection}: placing on the originating venue
  of each shadowed add makes the execution venue distribution match the observed
  add distribution without any venue score or routing optimizer, an inherited
  split rather than an optimized one (Section~\ref{sec:sor}). The evaluation
  corpus is a single venue, so this mechanism is unexercised by the measurements
  reported here; testing it requires a fragmented multi-venue market
  (Section~\ref{sec:future}).
  \item We describe the \emph{reference implementation} in the
  \textsc{kaspar-hft} system, which layers state-based
  safety guards (a two-band place/cancel hysteresis plus completion and
  over-hedge guards) atop the core method, distinguishing the minimal core from
  the implemented superset (Section~\ref{sec:prod}).
  \item We report a measured evaluation on a full calendar year of ES futures,
  run through the \textsc{kaspar-hft} deterministic replay simulator, in which
  the execution style and the simulated latency are varied one at a time and
  every other parameter is held fixed (Section~\ref{sec:eval}).
\end{enumerate}

\section{Prior art: existing execution methods}
\label{sec:priorart}

Automated execution in current practice is organized around two broad families:
schedule-based methods and liquidity-seeking methods
\cite{johnson2010,kissell,laruelle2011}. We describe each operationally and note how \pov{}
differs; the academic results underpinning these methods are reviewed separately
in Section~\ref{sec:lit}, and the closest patented mechanisms in
Section~\ref{sec:patents}.

\paragraph{Scheduled slicing (TWAP, VWAP).} The most widely deployed methods
slice a parent order onto a schedule. TWAP spreads child orders evenly in
time. VWAP shapes the schedule to a forecast of the intraday volume curve so
that the realized average price tracks the volume-weighted benchmark. Both
commit, before trading, to a trajectory keyed to a clock or a predicted volume
profile, and both must post ahead of where volume will be in order to
participate. \pov{} commits to no such trajectory: it places only in reaction to
orders already resting in the book, and does not forecast or pre-position against
future volume. Kissell \cite{kissell} covers both TWAP and VWAP in detail.

\paragraph{Percentage-of-volume (POV).} A POV algorithm holds a target
participation rate in market volume, scaling its trading up and down with
\emph{realized} volume rather than a fixed clock. It does not forecast a volume
curve, but it does track an \emph{aggregate} statistic (total traded volume)
and, to participate passively, must lean slightly ahead of near-term volume
rather than cross the spread. \pov{} is POV-like in effect --- its trading scales with the rate of observed
adds --- but its participation emerges from following \emph{individual} resting
orders at a bounded rate, and it does not pre-position against future volume. Kissell
\cite{kissell} covers POV in detail.

\paragraph{The POV control loop, and how \pov{} differs.} A conventional POV
algorithm runs a feedback loop against an aggregate statistic. Let $\rho$ be the
target participation rate, $V_t$ the cumulative market volume, and $x_t$ the
quantity executed so far. At each step the algorithm compares $x_t$ against the
target $\rho V_t$: when $x_t < \rho V_t$ it trades more, posting aggressively or
crossing the spread to catch up; when $x_t > \rho V_t$ it slows or pauses.

\pov{} replaces this control loop with per-order following
(Section~\ref{sec:algo}). Its actions are driven by the arrival of individual
orders: on each observed add it may post a shadow at that order's price and
venue, bounded by that order's size. The shadow rests just behind a live order at
the same price, so under FIFO priority a sweep must consume the followed order
and the depth ahead of the shadow before reaching it. Participation emerges from the rate of
order arrivals, placement and venue selection are inherited from the followed
order, and a forecast of forward volume is unnecessary. Table~\ref{tab:pov}
contrasts the two mechanically.

\begin{table}[h]
\centering
\caption{Conventional POV and \pov{}, by mechanism.}
\label{tab:pov}
\small
\begin{tabular}{p{3.3cm}p{4.7cm}p{4.7cm}}
\toprule
Aspect & Conventional POV & \pov{} \\
\midrule
Quantity tracked & aggregate market volume $V_t$ & individual resting orders \\
\addlinespace
Participation & target rate $\rho$, held by a feedback loop & emergent from following adds at a bounded rate \\
\addlinespace
Passive placement & leans ahead of forecast volume & mirrors orders already resting \\
\addlinespace
When behind & crosses the spread, paying impact & following continues until the target size is filled \\
\addlinespace
Book / venue placement & separate SOR and queue logic & inherited from the followed order \\
\addlinespace
Fill rate & depends on placement quality & requires a sweep that clears the followed order first (shadow rests just behind) \\
\bottomrule
\end{tabular}
\end{table}

\paragraph{Smart order routing (SOR).} When liquidity is fragmented across
venues, SOR decides where to send each child order by computing and comparing
per-venue liquidity, fees and rebates \cite{battalio2016}, fill rates, and
latency, and optimizing a routing objective
\cite{foucault2008,laruelle2011,contkukanov2015}. \pov{} performs none of this: it sends each shadow to the
originating venue of the add it follows, so the venue mix is inherited from
observed order flow (Section~\ref{sec:sor}) rather than solved for.

\paragraph{Liquidity-seeking algorithms.} A liquidity-seeking algorithm aims to
complete an order by capturing whatever liquidity is available at acceptable
prices, prioritizing completion and low information leakage over any schedule or
benchmark. As the second family of execution methods, it abandons a fixed schedule
and sources liquidity opportunistically: posting and taking across lit and dark
venues, and using hidden and iceberg orders and immediate-or-cancel probes to
find undisplayed block size \cite{kissell,laruelle2011}. It raises
participation when liquidity appears at an acceptable price and withdraws
otherwise, and may finish early when a block is found or wait when nothing
suitable appears. \pov{} operates on CLOB
venues only, following the \emph{displayed} resting orders on the book. It does
not route to dark pools and makes no attempt to detect hidden or iceberg
liquidity --- it follows what the book shows, and commits to no trajectory of
its own. \pov{} and a liquidity seeker pursue different goals: a liquidity
seeker minimizes footprint and hunts for size across lit and
dark venues; \pov{} places a passive order well on the lit book once a schedule
has already decided to rest one. The two families are complementary --- a
liquidity seeker decides \emph{when} enough liquidity is present
to act, whereas \pov{} decides \emph{where} to rest a passive order.

\paragraph{Quote matching.} The closest prior strategy in spirit is quote
matching, described by Harris \cite{harris2003}: a trader observes a large
standing limit order and places an order one tick \emph{better}, taking priority
ahead of it. The standing order is thereby converted into a free option --- if
the price moves against the quote matcher, it exits by trading \emph{against}
that order, which is the very liquidity it stepped in front of. This shares one
element with \pov{}: both act on an \emph{individual} identified resting order
rather than on aggregate book state. Everything else is opposed.

\pov{} places at the \emph{same} price, joining behind the observed order in the
queue rather than ahead of it, so it takes no priority the other participant
would otherwise have held. It shadows at a fixed rate irrespective of size,
rather than selecting large orders for the size of the option they offer. It
never trades against the order it follows: when that order ends, the shadow is
withdrawn (Section~\ref{sec:cancel}). And its purpose is to complete a parent
order at a predictable cost, not to extract an option from another participant's
commitment.

The comparison is also where the cancellation mechanism comes from. Harris's
stated objection to quote matching is that the large standing order can simply
vanish, leaving the quote matcher exposed precisely when the price then moves
against it. \pov{}'s exposure to that event is the same in kind, and
identifier-driven cancel mirroring is the direct answer to it: the shadow is
pulled on the delete or trade message for the specific order it follows, without
evaluating any market state (Sections~\ref{sec:map}--\ref{sec:cancel}). The
failure mode that makes quote matching hazardous is the event this method is
built to react to.

\section{Related literature}
\label{sec:lit}

The methods above rest on, and are analyzed by, a substantial academic
literature; we review the strands most relevant to \pov{}'s design choices.

\paragraph{Optimal-execution theory.} The modern foundation is Almgren--Chriss
\cite{almgren2000}, which casts execution as minimizing a mean--variance
trade-off between market impact (favoring slow trading) and timing/volatility
risk (favoring fast trading), yielding an optimal schedule given models of
permanent and temporary impact. A large body of work refines this framework,
including limit-order formulations with adverse selection and fill uncertainty
in the spirit of Avellaneda--Stoikov \cite{avellaneda2008}. This theory
answers \emph{how fast} to trade given an impact model and a volume profile;
it presupposes the profile. \pov{} keeps the passivity these analyses target
but does not compute a trajectory --- it follows one the market reveals.

\paragraph{Queue position and adverse selection.} Under price--time priority,
both fill probability and adverse-selection cost depend on where a resting
order sits in the first-in, first-out (FIFO) queue. Moallemi--Yuan \cite{moallemi2016} give a
valuation model in which adverse-selection cost \emph{increases} with queue
position, with a dynamic component capturing the option value of locking in a
position early. This literature quantifies precisely the effect \pov{} is
designed around: leading the book --- posting early and deep in a thin queue
--- buys priority but courts selection when the level thins ahead of an adverse
print. By posting alongside an order a participant has already chosen to
rest, \pov{} avoids deliberately taking priority ahead of that order. The
followed order therefore provides an observable reference for the local queue
environment without requiring \pov{} to predict that environment.

\paragraph{Order-flow competition and routing.} Foucault--Menkveld
\cite{foucault2008} analyze competition for order flow across venues and the
routing systems that arbitrate it, and Laruelle--Lehalle--Pag\`es
\cite{laruelle2011} give stochastic-approximation algorithms that learn the
optimal split of an order across liquidity pools. Where this literature
\emph{solves} for the venue split, \pov{}'s split \emph{emerges} from placing
on the originating venue of each followed order (Section~\ref{sec:sor}).

\paragraph{Practitioner references.} The practical construction and calibration
of VWAP/TWAP/POV and their trade-offs are documented in practitioner texts such
as Kissell \cite{kissell}, which we take as the reference description of the
methods in Section~\ref{sec:priorart}.

\paragraph{Recent execution research.} Contemporary execution research is
largely forecast- and optimization-based, which sharpens the contrast with
\pov{}. Reinforcement-learning formulations place market and limit orders to
minimize implementation shortfall, both in general limit-order settings
\cite{rlexec2025} and within queue-reactive order-book models
\cite{queuereactive2025}, and signal-adaptive schemes optimize quote placement
from predictive features. Optimal execution under \emph{passive} market impact
\cite{passiveimpact2026} models fill and impact as functions of spread, queue
position, displayed depth, imbalance, and recent order flow --- the same
state \pov{} does not compute. Work on the market maker's trade-off between
fill probability and post-fill return \cite{mmdilemma2025} quantifies the
top-of-book adverse selection that any resting order, including a shadow, inherits
(Section~\ref{sec:disc}). \pov{} differs from all of these in
computing nothing at all: it places in reaction to observed resting orders.

\section{Prior patents}
\label{sec:patents}

Four patented mechanisms sit closest to \pov{}. Each shares a surface
resemblance --- passivity, pegging, or following --- but differs in a way that
is material to \pov{}'s design.

\begin{itemize}[leftmargin=1.4em]
  \item \textbf{Pegged / tracking orders} (e.g.\ US~10,614,520,
  ``Tracking Liquidity Order'' \cite{us10614520}). A non-displayed order pegs
  to the same side of the National Best Bid and Offer (NBBO), continuously
  \emph{re-pricing against evaluated market state}. \pov{} evaluates no such state to manage its shadow: price is
  inherited once from the shadowed add, and the shadow is \emph{cancelled} when
  that add ends.
  \item \textbf{Copy / mirror trading} (e.g.\ US~2013/0268423, ``Copy Trading
  System and Method'' \cite{us20130268423}). These replicate \emph{executed
  trades} of \emph{identified} accounts a user has chosen to follow. \pov{}
  follows \emph{anonymous passive adds} at the message level --- the posting of
  liquidity itself --- and couples both entry \emph{and cancellation} to the
  followed order's identifier.
  \item \textbf{Exchange-side passive/hidden liquidity} (e.g.\
  US~2006/0253379, ``Passive Liquidity Order''~\cite{us20060253379}). This is a matching-engine
  mechanism internal to a venue; \pov{} instead is a client-side strategy that
  observes a public feed and places and cancels its own orders.
  \item \textbf{Hidden-liquidity detection} (e.g.\ US~8,140,416~\cite{us8140416}). Here order
  identifiers are tracked for \emph{analysis} (inferring hidden size), and
  trading decisions still evaluate market state; \pov{} uses identifier
  presence directly as the trigger for a cancellation.
\end{itemize}

\section{Positioning relative to prior work}
\label{sec:novelty}

Having reviewed the existing execution methods (Section~\ref{sec:priorart}),
the academic literature (Section~\ref{sec:lit}), and the closest patented
mechanisms (Section~\ref{sec:patents}), we state where \pov{} sits relative to
all three. The contribution is not the general idea of following displayed
liquidity or using participation rates, which are established in both the public
literature and prior patents. It is the specific combination of order-level
shadowing, identifier-keyed lifecycle mirroring, and same-venue placement as a
model-free passive-placement benchmark --- a combination of design choices that
all follow from one commitment, to build no model of the book and follow
observed orders instead:

\begin{enumerate}[leftmargin=1.6em]
  \item \textbf{Model-free placement.} The passive order's price is inherited
  from the observed add it follows, so placement needs no order-book model and no
  fill-probability prediction; this is what lets \pov{} serve as a benchmark for
  predictive placement models (Section~\ref{sec:disc}).
  \item \textbf{Identifier-keyed cancellation.} A shadow-order is withdrawn when a
  delete or trade message for the specific order it follows arrives, rather
  than when evaluated market state (spread, queue decay, fill probability)
  crosses a threshold.
  \item \textbf{Flow-inherited venue selection.} Each shadow is sent to the
  originating venue of the add it follows, so venue choice is a pass-through
  rather than an optimization; the cross-venue mix of executions thereby tracks,
  in expectation, the cross-venue mix of observed liquidity provision without
  explicit routing logic --- the inverse of the SOR approach
  \cite{foucault2008,laruelle2011}, where the split is solved for.
\end{enumerate}

Table~\ref{tab:novelty} summarizes the positioning.

\begin{table}[h]
\centering
\caption{How \pov{} differs from the nearest prior work.}
\label{tab:novelty}
\small
\begin{tabular}{p{5.0cm}p{4.3cm}p{5.4cm}}
\toprule
Prior work & Mechanism & How \pov{} differs \\
\midrule
Almgren--Chriss \cite{almgren2000} &
Optimal \emph{schedule} from a volume/impact model &
No schedule; participation emerges from following individual adds \\
\addlinespace
POV \cite{kissell} &
Reactive participation to \emph{aggregate} realized volume; leans ahead to
stay passive &
Follows \emph{individual} resting orders; never posts ahead \\
\addlinespace
Moallemi--Yuan queue value \cite{moallemi2016} &
Quantifies queue-position adverse selection &
Does not lead the queue; inherits a resting order's neighborhood \\
\addlinespace
SOR \cite{foucault2008,laruelle2011} &
Computes or learns the venue split &
Split \emph{emerges} from same-venue placement; no routing computed \\
\addlinespace
Tracking/pegged orders \cite{us10614520} &
Re-prices against evaluated NBBO state &
No state evaluated; cancels on followed-order termination \\
\addlinespace
Copy/mirror trading \cite{us20130268423} &
Copies executed trades of identified accounts &
Follows anonymous passive \emph{adds}; keys cancellation to their identifier \\
\bottomrule
\end{tabular}
\end{table}

\section{The \pov{} Method}
\label{sec:algo}

\subsection{Message model}

\pov{} consumes a real-time stream of order-book messages from one or more
venues (e.g.\ an ITCH-style binary feed). Each message carries a
\emph{message type} --- \textsc{add} (a new resting limit order),
\textsc{delete} (cancellation), \textsc{trade} (execution), \textsc{modify}
(amendment) --- an \emph{exchange-assigned order identifier}, and an
\emph{originating venue identifier}, together with side/price/quantity for
\textsc{add}/\textsc{modify}. Individual order lifecycles are therefore
followable at the message level by identifier.

\subsection{Core mechanic: participant shadowing by identifier}

On each \textsc{add} that satisfies a selection criterion (below), \pov{}
transmits a \emph{shadow} limit order at the add's price to the add's
originating venue, and receives a venue-assigned shadow-order identifier. It
then records a single association keyed by the observed order's
$(\text{venue},\texttt{exchange\_order\_id})$ --- exchange order identifiers are
unique only within a venue --- and mapping it to its shadow-order identifier,
\[
  (\text{venue},\;\texttt{exchange\_order\_id}) \;\longmapsto\; \texttt{shadow\_order\_id}.
\]
The shadow is thereafter \emph{bound} to the
observed order by that key: it inherits that order's price and venue, and ---
crucially --- its \emph{lifecycle}. The single invariant that makes this
``shadowing'' rather than ``quoting'' is this attachment; a shadow is a
projection of one identified resting order.

\subsection{Order-placement rate}
\label{sec:select}

\pov{} shadows a bounded fraction of qualifying adds. The reference
implementation offers two independent selection modes:
\begin{itemize}[leftmargin=1.4em]
  \item \emph{Probabilistic}: shadow a qualifying add with a fixed probability
  $\rho$, drawn independently per light. $\rho$ is a free parameter; the values
  used here are stated with each experiment.
  \item \emph{Deterministic}: shadow every $N$-th qualifying add.
\end{itemize}
The two modes give different selection rates. The experiments reported in
Section~\ref{sec:eval} use the probabilistic mode.
The quantity \pov{} sets is thus the \emph{order-placement rate} $\rho$ --- the
fraction of qualifying adds it shadows --- and \emph{not} a participation rate.
This placement rate is fixed by construction and applied to observed order flow,
not to traded volume; \pov{} places more shadows when adds are frequent and
fewer when they are sparse. The realized \emph{participation} --- \pov{}'s share
of traded volume --- cannot be controlled directly; it emerges from following at
the fixed placement rate. Because not every placed shadow fills, the realized
participation is not equal to the order-placement rate and will generally differ
from it.
Participation is therefore a random quantity whose dependence on parent-order
size is analyzed in Section~\ref{sec:partsize}.

\subsection{Shadow quantity}

The shadow's \emph{desired} quantity is computed independently of the triggering
add's quantity --- in the reference implementation as a randomized function of
average trade sizes over a preceding window. Before transmission the desired
quantity is capped by the followed order's size, the per-order and per-level
limits (e.g.\ 1-lot child orders with a 25-lot aggregate ceiling), and the
remaining parent quantity. Fills
follow from co-resting at the same price: a sufficiently large sweep can consume
the followed order and continue into the shadow, while a partial sweep may stop
before reaching the shadow.

\subsection{Retained information}
\label{sec:map}

For each followed order, the core association retains \emph{only} its
$(\text{venue},\;\text{exchange identifier})$ key paired with the shadow
identifier; it
does not retain the followed order's price, quantity, queue position, order-book
state, or timestamp.
The algorithm additionally maintains global configuration --- target side,
selection rate, and grace-window parameters --- together with the minimal state
needed to manage its own orders. Two consequences follow. First, no
price/queue/state-based
cancellation decision is \emph{possible}, because the inputs to such a
decision are not retained; cancellation is forced to be identifier-driven by
construction. Second, storing, looking up, and removing an association by
identifier is expected $O(1)$, so the cost of the cancellation decision is
independent of the number of tracked orders.

\subsection{Identifier-driven cancel mirroring}
\label{sec:cancel}

On a lifecycle-terminating message for a followed order, \pov{} looks up the
message's exchange identifier. If no association is present, nothing happens.
If it is present, the response depends on \emph{how} the followed order ended:

\begin{itemize}[leftmargin=1.4em]
  \item \textbf{Pulled (\textsc{delete}).} The liquidity simply left; no
  aggression is in flight to fill against. \pov{} cancels the shadow
  \emph{immediately} and removes the association --- continuing to rest there
  would be pure adverse-selection exposure.
  \item \textbf{Hit (\textsc{trade}).} A trade is a marketable sweep still in
  flight, and because the shadow rests in the same queue behind the followed
  order, that sweep may be about to reach it. Cancelling instantly would pull
  the shadow out of the queue \emph{before} the aggression arrives --- and in
  the worst case the shadow would then \emph{never} fill, because it always
  cancels in time. \pov{} therefore arms a \emph{delayed cancel}: it holds the
  shadow live for a short grace window of $B$ end-of-burst events, during which
  the shadow may itself be filled by the continuing sweep. If the window
  elapses without a fill, the cancel is sent and the association removed. Arming
  is idempotent: a further trade for the same order while the window is armed
  does not restart it.
\end{itemize}

Aside from this trade-versus-pull
distinction, the cancellation input is still just identifier presence: no book
depth, queue position, price competitiveness, or fill probability is evaluated.
The hard safety guards of the reference implementation
(Section~\ref{sec:prod}) --- target reached, over-hedge, or price drifted past
the cancel band --- may fire an \emph{immediate} cancel that pre-empts an armed
grace window. A shadow therefore never outlives the order it follows by more
than the grace window, preserving the discipline that \pov{} rests only where a
currently-live participant order rests, or rested an instant ago. A
\textsc{modify} for a tracked order is treated as termination of the followed
lifecycle: the shadow is cancelled and the association removed. Re-shadowing is
not automatic --- the amended order is shadowed anew only if the feed subsequently
emits a fresh qualifying \textsc{add} for it. This is an algorithmic convention
rather than an assertion about exchange modify semantics.

\subsection{Flow-inherited venue selection}
\label{sec:sor}

Because each shadow is transmitted to the originating venue of the add it
follows, a fixed-rate selector preserves, in expectation, the cross-venue
distribution of qualifying observed adds. If participants rest 50\%/30\%/20\%
of adds on venues A/B/C, a fixed selection rate reproduces that 50:30:20 mix
in expectation --- \emph{without} venue liquidity metrics, fee/rebate
optimization, latency-based routing, or any explicit routing logic. This is
flow-inherited venue selection, not an optimization: the venue mix is
proportional to the observed distribution of qualifying adds, which need not
coincide with the split that minimizes fees, latency, or adverse selection, nor
with the distribution of traded volume.

\begin{algorithm}[t]
\caption{\pov{} identifier-driven core (per message)}
\label{alg:core}
\DontPrintSemicolon
\SetKwInOut{State}{state}
\State{mapping $M:\;k\!\mapsto\!\texttt{shadow\_id}$ with venue-scoped key
       $k=(\text{venue},\texttt{ex\_id})$ (exchange ids are unique only within a
       venue); target side $\sigma$; order-placement rate $\rho$; grace window
       $B$ bursts; own-id set $O$ of $(\text{venue},\texttt{shadow\_id})$ pairs}
\BlankLine
\KwIn{message $m$ with type $m.\tau$, exchange id $m.\text{id}$, venue
      $m.v$, side $m.s$, price $m.p$; write $k=(m.v,\,m.\text{id})$}
\BlankLine
\Switch{$m.\tau$}{
  \uCase{\textsc{add}}{
     \lIf{$k\in O$ \textbf{or} $m.s\neq\sigma$}{\Return \tcp*[f]{self / wrong side}}
     \lIf{selection criterion not met (prob.\ $\rho$ / every-$N$)}{\Return}
     $\texttt{shadow\_id} \gets$ transmit shadow at price $m.p$ to venue $m.v$;\;
     $O \gets O\cup\{(m.v,\texttt{shadow\_id})\}$;\quad $M[k] \gets \texttt{shadow\_id}$\;
  }
  \uCase{\textsc{delete} \textup{(followed order pulled)}}{
     \lIf{$k\in M$}{cancel $M[k]$ now; remove $M[k]$}
  }
  \uCase{\textsc{trade} \textup{(followed order hit)}}{
     \If{$k\in M$ \textbf{and} not already armed}{arm delayed cancel of $M[k]$ in $B$ bursts
        \tcp*{grace: the sweep may still fill the shadow}}
  }
  \Case{\textsc{modify}}{
     \If{$k\in M$}{cancel $M[k]$ now; remove $M[k]$ \tcp*[f]{teardown}}
  }
}
\BlankLine
\tcp{grace-window expiry or safety guard $\Rightarrow$ cancel armed shadow; own fill $\Rightarrow$ remove its association}
\tcp{on parent complete: cancel all shadows in $M$; clear $M$ (wind-down)}
\end{algorithm}

\section{Reference Implementation: the \textsc{kaspar-hft} System}
\label{sec:prod}

\textsc{kaspar-hft} is an open-source, low-latency trading system and order book simulator written in
C++.

\subsection{Simulation facilities}
\label{sec:sim}

The simulator is the instrument this paper is
written with, so it is worth stating what it provides. It
reconstructs a full order-by-order (MBO) book from recorded exchange packet
captures, maintaining every resting order individually rather than aggregated
depth, which is what makes a queue position --- and therefore a fill inference
--- definable at all. Three latencies are configurable and independent: the
outbound order, the outbound cancel, and the inbound market-data feed.

\paragraph{The fill inference.}
A shadow enters the simulated price-time-priority (FIFO) queue at its price, behind
the depth already resting there. It is assigned a fill when subsequent recorded
executions imply that aggression has reached a resting order \emph{behind} it:
the shadow has earlier time priority than that order, so an aggressor that
reached the one must have passed through the other.

\paragraph{Market impact.} The simulator leaves it out. Orders are filled
against the recorded feed as though they had not been there: the book carries on
exactly as it did on the day, so a quote the algorithm sits in front of stays
where it is and a counterparty that would have repriced does not. Every cost
reported here is therefore a cost before the market's reaction to the order that
produced it.

It is built using the actor-model architecture
\cite{hewitt1973,agha1986}: independent components communicate
only by passing messages, so concurrent order handling stays deterministic and
keeps shared-memory locking off the critical path. The same code runs unchanged
across backtesting, market replay, live simulation, and live trading; the
evaluation of Section~\ref{sec:eval} therefore exercises the live code path
directly.

\subsection{Passive POV implementation}

\pov{} is implemented as a set of execution components within
\textsc{kaspar-hft}. A strategy issues a position order, and the execution
components translate it into the individual limit orders sent to the exchange,
with the acquiring and liquidating sides of each instrument handled
independently. This implementation is a \emph{superset} of
Algorithm~\ref{alg:core}: it keeps the identifier-driven attach/cancel core and
adds state-based \emph{safety guards} appropriate to running real capital
against a live venue.

The added guards are:

\begin{itemize}[leftmargin=1.4em]
  \item \textbf{Two-band place/cancel hysteresis.} A shadow is only
  \emph{placed} when its price is within a tight placement band
  $\delta_{\text{place}}$ ticks of the inside (e.g.\ for a buy,
  $p \ge b_0-\delta_{\text{place}}$ against best bid $b_0$), but is only
  \emph{cancelled for distance} once it drifts beyond a looser cancel band
  $\delta_{\text{cancel}}$, with $\delta_{\text{place}}<\delta_{\text{cancel}}$.
  The gap introduces hysteresis: it
  prevents a one-tick oscillation of the inside from triggering a
  place/cancel/replace loop that would burn message budget and queue priority.
  The placement band $\delta_{\text{place}}$ also bounds how far from the inside a
  shadow can attach; tightening it restricts shadowing to adds near the touch,
  which fill more reliably and reduce the small-order participation dispersion of
  Section~\ref{sec:partsize}.
  \item \textbf{Completion and over-hedge guards.} A working shadow is also
  pulled when the position reaches its target or goes flat, when the size
  resting at its price would overshoot the quantity that remains to be executed,
  or when per-level order caps are exceeded. The placed size is the minimum of
  the per-level cap, the per-order cap, the remaining quantity, and the followed
  order's size, clamped at zero. Between them the guards hold the shadow within
  the quantity that remains and within the size of the participant it follows.
\end{itemize}

\paragraph{Grace window.} As described in Section~\ref{sec:cancel}, the shadow's
cancel is delayed only when the followed order is hit by a trade, to give the
incoming sweep a chance to fill the shadow; when the followed order is instead
cancelled, the shadow is cancelled at once. The safety guards override this
delay: whenever a completion, over-hedge, or distance limit is reached, the
shadow is cancelled immediately.

In short, the core method is the minimal identifier-driven algorithm
(Algorithm~\ref{alg:core}); the reference implementation runs that core plus the
guards above. Guarded cancellations are a strict addition to the
identifier-keyed cancel.

The guards read market state --- the inside price and recent trade sizes --- and
it is worth being precise about what that does and does not mean for the
model-free claim. \emph{Model-free} here means what it means throughout the
paper: no order-book model and no fill prediction. A band on distance from the
touch, a cap on size resting at a price, and a limit on position against target
are risk limits. None of them forecasts a fill, values a queue position, or
estimates where the price is going, and none of them chooses \emph{where} to
place --- the price is still inherited from the shadowed order, and the
cancellation is still driven by that order's identifier. Reading the inside
price to enforce a band is not a model of the book any more than a position
limit is a model of the market. The guards bound what the method may do; they
do not supply it with a view.

\subsection{Naive aggressive POV}
\label{sec:aggressive}

The aggressive baseline that Section~\ref{sec:eval} measures against is the same
shadowing mechanism driven by the \emph{opposite} signal. Where passive \pov{}
shadows resting \textsc{add}s, the aggressive arm shadows \emph{aggressive}
orders --- the trade prints. On each observed \textsc{trade}, it flips a
coin with a fixed per-trade probability (a placement rate in basis points of
print) and, when selected, places a marketable order
at the print's price that follows the aggressor into the book. This is a
\emph{naive} POV: it participates by taking a fixed fraction of trading activity
directly, with no volume-curve forecast and no participation-feedback loop ---
the same construction as passive \pov{}, applied to trades rather than adds.
In the reference implementation both arms are in fact the \emph{same} C++ class (Actor).

\paragraph{This arm is standard practice, not a construction of ours.} Print
following is what commercial POV implementations already do, and we claim no
novelty for it. Deltix's PVOL states the rule explicitly: it divides the order
duration into intervals, and ``at the end of each interval, PVOL calculates the
total volume of trades that happened on the market'', then sizes the child order
as $\text{Quantity} = \text{VOL}(T_i) \times \text{ParticipationVolume}$
\cite{deltixpvol} --- observe the prints, then send a marketable child sized as a
fraction of them. TradeStation describes its POV route the same way, as an order
that ``trades based on real time market volumes'' and is worked with market or
marketable limit orders at algorithm-determined intervals \cite{tspov}. A POV
algorithm is therefore a liquidity \emph{taker} that waits for volume to print
and participates afterwards \cite{kissell}.

The aggressive arm here is that rule taken to its limiting grain: instead of
aggregating prints over a 30-second interval and crossing once, it acts on each
print individually. That is a difference of granularity, not of kind, which is
precisely why it makes a fair baseline --- the passive method is being compared
against what desks actually run, rather than against a strawman built for the
purpose. The novelty claimed in this paper is on the \emph{passive} side
(Sections~\ref{sec:map}--\ref{sec:cancel}), where we have found no published
method that places at an identified individual order's price and ties
cancellation to that order's exchange identifier.

Two mechanical details distinguish it from the passive arm. The
aggressive order is an immediate-or-cancel in intent: it is given a short
time-to-live (default $1$~ms) so that a cross which arrives after the level it
was priced from has already been consumed is pulled rather than left to rest and
silently become a passive order the arm never asked for. This is
what is referred to as ``Aggressive POV'' throughout Section~\ref{sec:eval}.

We carry the aggressive arm because the comparison it enables is the practically
important one. The question a desk faces is not whether shadow-following works in
the abstract but how passive placement compares to simply crossing, and a
practitioner will typically run both versions side by side and choose per order.
Implementing the two in one class lets that comparison be made like-for-like ---
same instrument, corpus, and position mandate --- which is exactly the
passive-against-aggressive experiment of Section~\ref{sec:eval:results}.

\section{Experimental Evaluation}
\label{sec:eval}


We evaluate the baseline algorithm performance here. Since our simulator does not
model market impact, the algorithm's measured cost carries no component that
scales with participation rate or order size. We therefore focus on the base
case, where slippage is measured in idealized conditions. We then compare the
passive \pov{} against its aggressive equivalent. Finally we check the latency
degradation characteristics.

\subsection{What the experiment is}
\label{sec:eval:setup}

\paragraph{Instrument and corpus.} E-mini S\&P~500 futures (CME product ES),
market-by-order data on CME channel~310, every regular trading session of
calendar year 2025 for which PCAP files were available.

\paragraph{What the algorithm is asked to do.} The algorithm is run as a
\emph{two-sided market maker}, quoting the buy and the sell side simultaneously
throughout the regular session, and is required to complete a fixed quantity on
each side within each measurement window. Concretely:

\begin{itemize}[nosep,leftmargin=1.4em]
  \item \textbf{Window boundaries} are placed every 10 minutes from 09:29~ET to
        16:00~ET. Each boundary closes one window and opens the next, so a
        session yields at most 39 windows.
  \item \textbf{At each boundary} the algorithm is given 100 contracts to buy
        and, independently, 100 contracts to sell. The two are worked over the
        same window by separate books, so both sides quote at once rather than
        one after the other.
  \item \textbf{A window closes} when \emph{both} conditions hold: the 10-minute
        minimum has elapsed, \emph{and} both sides have completed their full 100
        contracts. A session in which the
        algorithm fills slowly therefore produces \emph{fewer and longer}
        windows, never a partially filled one.
\end{itemize}

\subsection{What is measured}
\label{sec:eval:measures}

All prices are in ticks; one tick is $0.25$ index points, which is
$\$12.50$ per contract on ES.

\paragraph{Relative slippage.} Let $\text{vwap}_{\textsc{buy}}$ and
$\text{vwap}_{\textsc{sel}}$ be the size-weighted average prices at which the
window's 100 bought and 100 sold contracts were actually executed, and let
$m_{0}$ be the mid-price at the instant the window opened --- the arrival price
for both legs, since both start together. Each leg is priced against that mid,
\begin{equation}
  \text{Slippage}_{\textsc{buy}} = \text{vwap}_{\textsc{buy}} - m_{0},
  \qquad
  \text{Slippage}_{\textsc{sel}} = m_{0} - \text{vwap}_{\textsc{sel}},
  \label{eq:slipleg}
\end{equation}
signed so that a positive value is a loss on either side, and the reported
figure is the average of the two:
\begin{equation}
  \text{RelativeSlippage} \;=\; \tfrac12\big(\text{Slippage}_{\textsc{buy}}
                                   + \text{Slippage}_{\textsc{sel}}\big)
                 \;=\; \tfrac12\big(\text{vwap}_{\textsc{buy}}
                                   - \text{vwap}_{\textsc{sel}}\big).
  \label{eq:slippage}
\end{equation}

We call this \emph{relative slippage}: relative to the arrival mid, and
relative between the legs. The second equality is the reason to average rather
than report the legs separately: $m_{0}$ cancels, so whatever the market did
during the window affects both legs and drops out. The per-leg figures are considerably noisier
for exactly that reason --- over a leg of a minute or more the market drifts, and
the drift lands in the two legs with opposite signs. Measured over the corpus the
buy and sell legs carry 95\,\% intervals about four times wider than their own
average. The paired form is immune to that drift by construction, which is what
makes a full-year average meaningful.

Both legs start together but do not necessarily finish together.

Quoting both sides at once raises the question of whether the algorithm's own
bid- and offer-side shadows interact. They cannot trade against each other: the
simulator matches each shadow against the recorded feed, never against another
shadow. Nor can they queue behind one another, since each attaches to a
third-party order on its own side of the book. On ES, which is one tick wide at almost every window
opening, there is rarely an intervening price level at which the two sides could
meet. The interaction that does remain is through the position
and completion guards of Section~\ref{sec:prod}, which see net position and can
therefore pull a shadow on one side because of fills on the other.

\paragraph{Cost relative to immediate execution.} The alternative to placing
passively is to cross the spread at once, priced against the same $m_{0}$: a buy
crosses to the ask and a sell to the bid, so immediate execution has a slippage
of half the quoted width. Over this corpus the ES book is one tick wide at
$96.3\,\%$ of window openings, giving a mean spread of $1.038$ ticks and an
immediate-execution slippage of $+0.519$ ticks per contract. The difference,
\begin{equation}
  \text{RelativeToImmediate} \;=\; \text{RelativeSlippage} - \tfrac12\big(
        \text{ask}_{0} - \text{bid}_{0}\big),
  \label{eq:vstouch}
\end{equation}
is negative when the passive execution finished at a better average price than
crossing at the open would have. Both quantities in it are slippages against the
same arrival mid, so the subtraction is between like things.

\paragraph{Cost relative to VWAP.} The same executions compared against the
volume-weighted average price of every trade that occurred in the market while
the leg was working, rather than against the arrival mid. A negative value
means the algorithm executed better than the average price paid by everyone else
trading over the same interval. This measure is not comparable across
configurations whose legs run for different lengths of time: each is scored
against its own interval, and a longer leg averages over a broader and more
forgiving slice of the day.

\paragraph{Participation rate.} For one side of one window, let $F$ be the
contracts the algorithm filled and $V$ the contracts traded by everyone else
over that same side's duration. The participation rate is
\begin{equation}
  \text{Participation} \;=\; \frac{F}{F+V},
  \label{eq:part}
\end{equation}
bounded in $[0,1]$, which is what a percentage-of-volume mandate means by the
term. One number is reported per cell: the \emph{aggregate} rate
$\sum F / \sum (F+V)$ over every side of every window, which is the share of the
year's volume the algorithm took and the figure a mandate is written against.
The mean of per-window ratios is not used --- it is dominated by windows with
very small denominators and reverses the ordering entirely. Sides alongside
which no other trade occurred leave the ratio undefined and are excluded rather
than recorded as 100\,\%.

Both sides of a window are pooled, because the window is a 100-contract buy and
a 100-contract sell and the two are symmetric by construction. Measured
separately they agree to within $0.08$ percentage points in every cell, so
pooling costs nothing and reporting one side alone would be an arbitrary choice
between two answers that do not differ.

\paragraph{Time to fill.} The mean wall-clock time for a side to complete its
100 contracts, pooling both sides. The 100 contracts arrive as roughly 85
separate executions of $1.18$ contracts on average, in every cell. The clip
size is not a parameter: a shadow's size is bounded by the size of
the order it follows, so it is the market's own add-size distribution that sets
it.

\subsection{Results}
\label{sec:eval:results}

\paragraph{Passive against aggressive.}
The informative comparison is between the two execution styles. Both execute the
same parent over the same corpus; one rests and one crosses; and they are
configured so that the realized participation is approximately equal, which is
the only basis on which their costs can be set side by side. Participation is not
a control input in either style: neither can be told to hit a target rate --- it
emerges from following at a fixed placement rate (Section~\ref{sec:select}) ---
so it cannot be forced to match. Instead of forcing it, we swept the placement
rates and selected the pair of settings whose realized participation came out
empirically closest, $4.46\,\%$ passive against $4.86\,\%$ aggressive.

\begin{table}[t]
\centering
\caption{Passive and aggressive execution at nearly matched participation. ES futures,
246 sessions of calendar 2025, 100 contracts bought and 100 sold in every
window, zero simulated latency on all three paths. Cost is in ticks per
contract: a positive slippage is a loss, a negative relative cost is a gain.
Intervals are 95\,\% and clustered by session, since windows within a day are
not independent. \emph{Crossed} is the share of executed quantity that filled on
arrival rather than resting: it is the line that separates the two styles
mechanically, and the cost lines agree across it.}
\label{tab:styles}
\small
\begin{tabular}{lrr}
\toprule
 & \pov{} (passive) & Aggressive POV \\
\midrule
Shadowed event              & resting add & trade print \\
Rate                        & 5\,\% of adds & 2\,\% of prints \\
Completed windows           & 9{,}592 & 9{,}581 \\
\midrule
Relative slippage (ticks/contract) & $+0.0955 \pm 0.0130$ & $+0.0952 \pm 0.0135$ \\
\quad of which, buy leg     & $+0.0704 \pm 0.0532$ & $+0.0292 \pm 0.0830$ \\
\quad of which, sell leg    & $+0.1205 \pm 0.0517$ & $+0.1613 \pm 0.0836$ \\
Relative to immediate execution & $-0.4236 \pm 0.0117$ & $-0.4239 \pm 0.0105$ \\
vs.\ VWAP                   & $-0.6208 \pm 0.0204$ & $-0.4036 \pm 0.0275$ \\
\midrule
Participation (\%)          & 4.46 & 4.86 \\
Time to fill 100 (s)        & 72.7 & 62.4 \\
Child executions per leg    & 85.5 & 85.6 \\
Crossed (\%)                & 1.04 & 51.86 \\
\bottomrule
\end{tabular}
\end{table}

Table~\ref{tab:styles} gives the relative slippage: $+0.0955$
ticks per contract for passive against $+0.0952$ for aggressive, and relative to
immediate execution $-0.4236$ against $-0.4239$ --- agreement to the fourth
decimal on both, over more than nine thousand completed windows each.

This is the prediction of the Glosten--Milgrom account of the
spread~\cite{glostenmilgrom1985} and is discussed in Section~\ref{sec:disc}: the
half-spread a resting order captures is returned, in expectation, through
adverse selection, and the half-spread a crossing order pays buys the avoidance
of it. A method with no forecast cannot separate the two, so the choice between
resting and crossing is a choice about speed and participation, not about price.

\paragraph{The VWAP column is the one that differs.}
At $-0.621$ against $-0.404$ passive \pov{} appears to beat the market by half
again as much. The difficulty is that each leg is scored against the
volume-weighted price over its own working interval, and those intervals are not
the same length --- in either style the buy leg and the sell leg finish at
different times, and passive \pov{}'s legs run longer than aggressive POV's. A
longer interval averages over a broader slice of the session. So even
where a leg beats VWAP, the margin does not translate into usable alpha, and the
column should not be read as a ranking of the two methods.

\paragraph{Post-fill mark-outs cross-check.} Table~\ref{tab:markout} reports the
mid-price move over the one, five and thirty seconds after each individual fill,
signed so that a positive value is a move against the fill. The mark-outs are
close to the relative-slippage measure, which is the result we expect. The two
are related rather than identical: one is the price a completed round trip
achieved, the other is how the mid moved after each individual fill, and they are
computed at different grains. They share the same fill inference, so agreement
does not verify that inference; what it does show is that the price a window
paid is consistent with what the market did immediately afterwards. A
discrepancy would indicate that the simulator is not pricing executions
correctly, and there is none.

\begin{figure}[!htbp]
\centering
\includegraphics[width=0.66\textwidth]{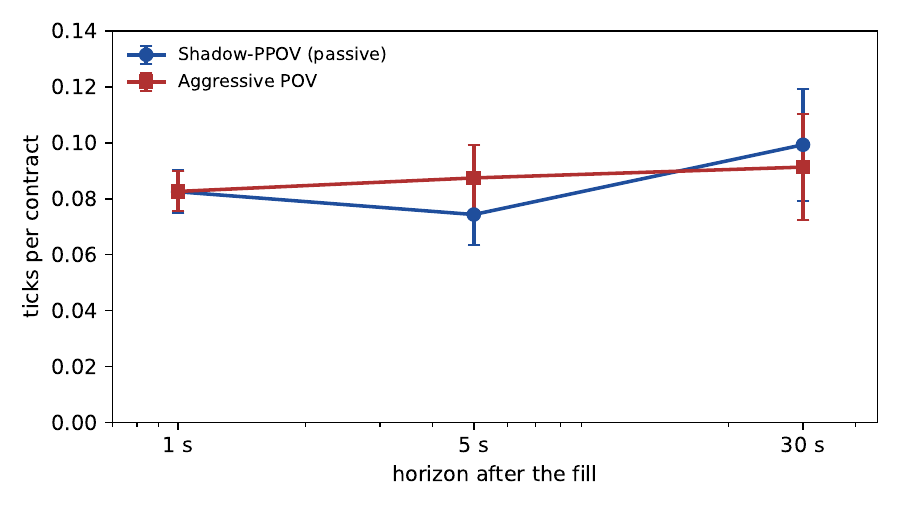}
\caption{Post-fill mark-out for the two execution styles, at one, five and
thirty seconds, with 95\,\% intervals clustered by session. The two are
indistinguishable at every horizon, and each profile is flat rather than
decaying: the price move is complete within a second of the fill and does not
revert over the following twenty-nine. That is the signature of adverse
selection rather than of temporary impact.}
\label{fig:markout}
\end{figure}

\begin{figure}[!htbp]
\centering
\includegraphics[width=0.48\textwidth]{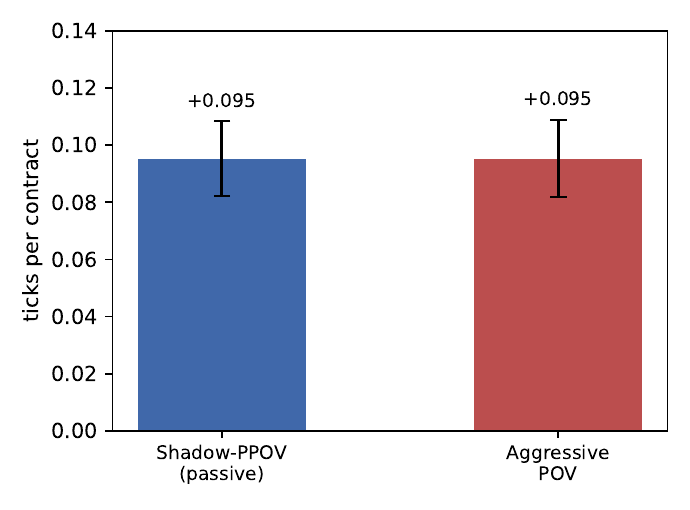}
\caption{Window-level relative slippage for the two execution styles at matched
participation, with 95\,\% intervals clustered by session. One style rests for
99\,\% of its executed quantity and the other crosses for half of it; the round
trip costs the same either way.}
\label{fig:slippage}
\end{figure}

\begin{figure}[!htbp]
\centering
\includegraphics[width=0.72\textwidth]{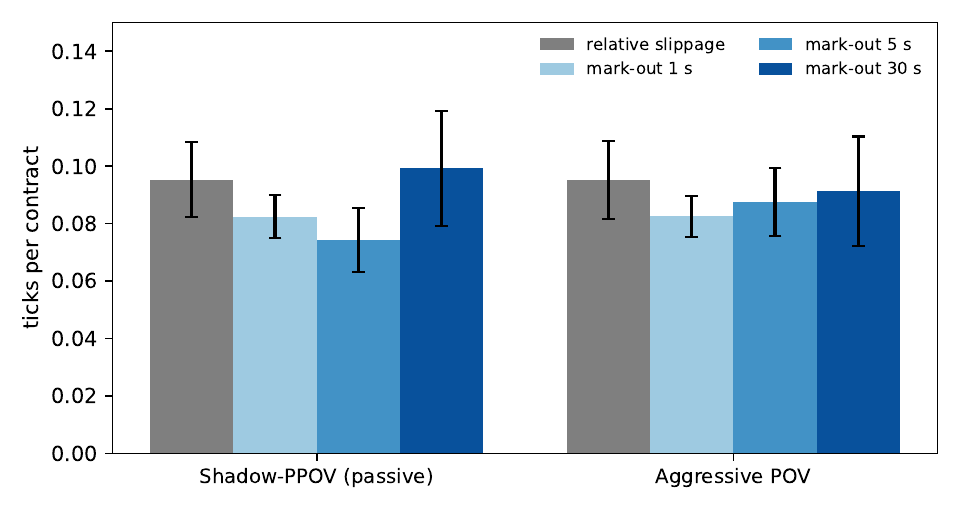}
\caption{Relative slippage against mark-out, both execution styles. The two quantities
are computed from different data at different grains --- one from the
volume-weighted prices of a whole window, the other from the mid-price after
each individual fill --- so their agreement is a check on the measurement rather
than an identity. They agree to within a couple of hundredths of a tick
throughout.}
\label{fig:crosscheck}
\end{figure}

Figure~\ref{fig:markout} compares the mark-out of the two execution styles,
Figure~\ref{fig:slippage} their relative slippage, and Figure~\ref{fig:crosscheck} puts
the two measures side by side for both.

\begin{table}[!htbp]
\centering
\caption{Three measures of the same executions, in ticks per contract.
\emph{Relative slippage} prices the two legs of a window against the
arrival mid and is a per-window quantity. \emph{Mark-out} is the mid-price move over the stated
horizon after each individual fill, one row per execution, signed so that
positive is a move against the fill. Both are priced against the mid, so they
are on the same scale and directly comparable. They measure \emph{related}
quantities rather than one quantity twice: relative slippage is the price a
completed round trip achieved, mark-out is how the mid moved after each
individual fill. Agreement between them is therefore a consistency check ---
what a window paid lines up with what happened after its fills --- and not a
replication.
Both rows cover the full corpus: 245 sessions and $1.63$ million individual
fills for the passive style, 246 sessions and $1.64$ million for the aggressive.
Intervals are 95\,\% and clustered by session.}
\label{tab:markout}
\footnotesize
\setlength{\tabcolsep}{4pt}
\begin{tabular}{lcccc}
\toprule
 & Rel.\ slippage & \multicolumn{3}{c}{Mark-out after the fill} \\
\cmidrule(lr){3-5}
 & (vs.\ mid) & 1\,s & 5\,s & 30\,s \\
\midrule
\pov{} (passive) & $+0.0953 \pm 0.0130$ & $+0.0825 \pm 0.0076$ & $+0.0743 \pm 0.0110$ & $+0.0993 \pm 0.0200$ \\
Aggressive POV   & $+0.0952 \pm 0.0135$ & $+0.0826 \pm 0.0072$ & $+0.0874 \pm 0.0118$ & $+0.0913 \pm 0.0190$ \\
\bottomrule
\end{tabular}
\end{table}

The mark-outs agree with the window-level costs and with each other. Both
styles give up something under a tenth of a tick per contract to the subsequent
mid-price, and the two are indistinguishable at every horizon. The profile is
flat in time rather than decaying: the price move is complete within a second of
the fill and does not revert over the following twenty-nine. That is the
signature of adverse selection rather than of temporary impact --- the fill
happened because someone with a reason to trade took the other side, and the
reason does not go away.

\FloatBarrier
\subsection{Latency}
\label{sec:latency}

Of everything varied in this study, the round-trip delay is the parameter that
moves the cost. It moves it monotonically, in both execution styles, with
non-overlapping confidence intervals from end to end.

The delay is applied to \emph{all three} paths at once --- the outbound order,
the outbound cancel, and the inbound market-data feed --- because a study that
delays only the outbound side measures an algorithm that still sees the book
instantly, which no participant does. Table~\ref{tab:latency} reports both
styles across the range.

\begin{table}[t]
\centering
\caption{Latency sensitivity, the same delay on order, cancel and feed. ES
futures, 246 sessions of 2025, 100 contracts bought and 100 sold per window,
every other parameter fixed within each style. The configurations here are
\emph{not} the pair of Table~\ref{tab:styles}: that pair was selected for the
closest realized participation match, whereas these two were each left at their
own working settings so that the delay is the only thing varied within a style.
Both styles therefore sit at a different placement rate here than in
Table~\ref{tab:styles}: the aggressive one shadows $1\,\%$ of prints against
$2\,\%$ there, which is why its zero-delay row reads $3.18\,\%$ participation
and $92$\,s rather than $4.86\,\%$ and $62.4$\,s, and the passive one reads
$4.51\,\%$ against $4.46\,\%$. Rows are comparable down a column; across
tables they are not.}
\label{tab:latency}
\small
\begin{tabular}{rrrrrr}
\toprule
Delay & Windows & Rel.\ slippage & vs.\ immediate & Participation & Time \\
($\mu$s) & & (ticks/contract) & (ticks/contract) & (\%) & (s) \\
\midrule
\multicolumn{6}{l}{\emph{Passive \pov{}}}\\
0 & 9{,}591 & $+0.097 \pm 0.013$ & $-0.422$ & 4.51 & 72 \\
500 & 9{,}593 & $+0.121 \pm 0.014$ & $-0.400$ & 6.00 & 55 \\
1000 & 9{,}592 & $+0.132 \pm 0.015$ & $-0.388$ & 6.12 & 54 \\
2500 & 9{,}593 & $+0.141 \pm 0.015$ & $-0.380$ & 6.40 & 52 \\
5000 & 9{,}594 & $+0.150 \pm 0.015$ & $-0.371$ & 6.87 & 48 \\
\midrule
\multicolumn{6}{l}{\emph{Aggressive POV}}\\
0 & 9{,}584 & $+0.087 \pm 0.017$ & $-0.432$ & 3.18 & 92 \\
500 & 9{,}541 & $+0.176 \pm 0.017$ & $-0.345$ & 2.43 & 121 \\
1000 & 9{,}525 & $+0.211 \pm 0.018$ & $-0.309$ & 2.24 & 131 \\
2500 & 9{,}531 & $+0.240 \pm 0.018$ & $-0.280$ & 2.09 & 138 \\
5000 & 9{,}564 & $+0.296 \pm 0.022$ & $-0.225$ & 2.02 & 141 \\
\bottomrule
\end{tabular}
\end{table}

\begin{figure}[!htbp]
\centering
\includegraphics[width=0.64\textwidth]{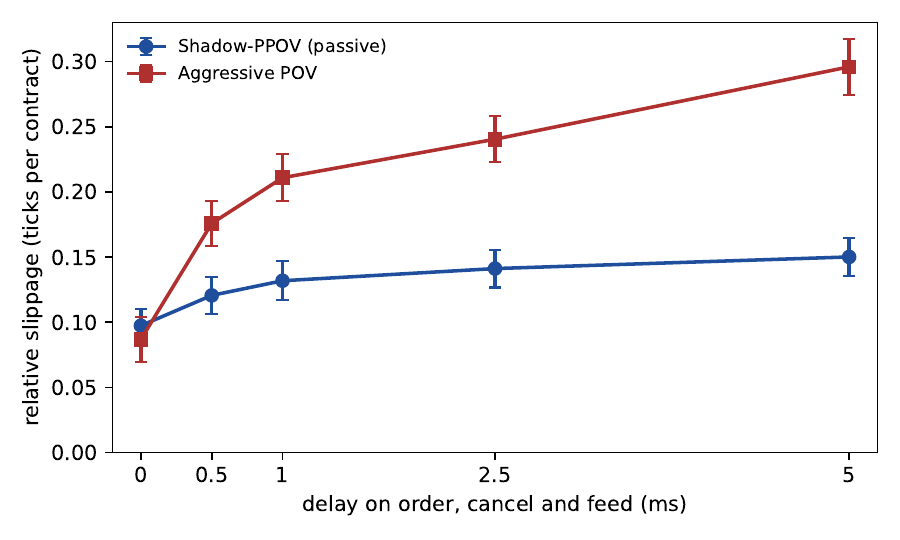}
\caption{Relative slippage against round-trip delay --- the cost column of
Table~\ref{tab:latency} --- with 95\,\% intervals clustered by session. The same
delay is applied to the outbound order, the outbound cancel and the inbound
feed. The two styles start together and separate at once: aggressive POV is the
cheaper of the two at zero delay and the more expensive by $500\,\mu$s, and its
slope over the range is about four times the passive one.}
\label{fig:latency}
\end{figure}

\paragraph{1. Cost rises monotonically.} Passive \pov{} costs
$+0.097$ ticks per contract at zero delay and $+0.150$ at 5\,ms, a 55\,\%
increase; aggressive POV goes from $+0.087$ to $+0.296$, more than a tripling.
Every step in both columns lies outside the preceding cell's interval, and
Figure~\ref{fig:latency} plots them. How fast the algorithm reaches the exchange
governs what it pays.

The two do not degrade at the same rate, and the difference is the more
interesting half of the table. Averaged over the range, passive \pov{} loses
$0.011$ ticks per contract per millisecond of delay and aggressive POV loses
$0.042$ --- roughly four times as fast. The ordering reverses inside the first
half-millisecond: at zero delay aggressive POV is the cheaper of the two by
$0.010$ ticks, and by $500\,\mu$s it is the more expensive by $0.055$. Measured
against immediate execution the asymmetry is starker still. Passive \pov{}
gives up $0.051$ of its $0.422$ advantage across the whole range, about an
eighth of it; aggressive POV gives up $0.207$ of $0.432$, very nearly half.
Everything concluded in Section~\ref{sec:eval:results} from the zero-latency
comparison --- that the two styles cost the same, and that the choice between
them is one about speed rather than price --- holds at zero latency and nowhere
else in this table.

\paragraph{2. Neither style holds its character.} The mechanism is the share of
executed quantity that fills on arrival rather than resting. A passive quote is
priced against a book that has moved by the time the order lands, so it arrives
marketable: passive \pov{} goes from $1.1\,\%$ of its quantity filling on
arrival to $27.3\,\%$, a quarter of it converted into liquidity taking by
latency alone, with no change to its logic. Aggressive POV
drifts the other way at first --- at 500\,$\mu$s its marketable limits arrive
after the level they were priced from has been consumed, and rest instead ---
and then upward, to $74.6\,\%$, as the market moves far enough during the flight
that the limit is through the touch on arrival. The two styles converge toward
each other.

\begin{figure}[!htbp]
\centering
\includegraphics[width=0.64\textwidth]{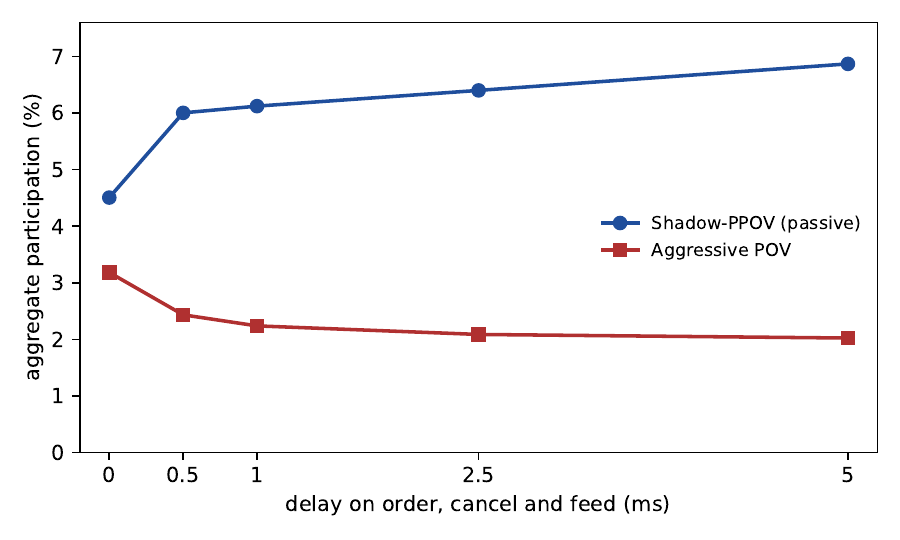}
\caption{Aggregate participation against round-trip delay, the same runs as
Figure~\ref{fig:latency}. Passive \pov{} takes a \emph{larger} share of the
market's volume as it gets slower and aggressive POV a smaller one. Both follow
from the conversion of execution style: the passive side turns quotes into
crosses and finishes sooner, the aggressive side loses crosses and takes longer,
and participation is scored over the leg's own duration.}
\label{fig:latencypart}
\end{figure}

\paragraph{3. Participation moves in opposite directions.}
Figure~\ref{fig:latencypart} is the one place in this study where the two styles
do not merely differ in degree. Passive participation \emph{rises} with delay,
$4.51\,\% \to 6.87\,\%$, half again as much of the market's volume; aggressive
participation \emph{falls}, $3.18\,\% \to 2.02\,\%$, to under two thirds of what
it was.

\paragraph{4. The degradation is in the price, not only in the mix.} Relative slippage
rises monotonically in both styles while the quantity executed and the number of
child fills are unchanged, so the algorithm is paying more for the same
executions rather than doing a different amount of business. How much of that is
the style conversion of item 2 and how much is worse execution within a style
cannot be separated from these runs alone; what can be said is that both
channels push the same way and neither reverses anywhere in the range.

\paragraph{5. The latency sensitivity is itself evidence that the method
extracts information from the flow.} A policy whose actions are independent of
order flow has no timing relationship with that flow, and therefore nothing for
delay to degrade: place at arbitrary times and prices and it should not much
matter whether the order reaches the exchange now or five milliseconds from now.
What Table~\ref{tab:latency} shows is the opposite. Cost rises monotonically in
both styles, with non-overlapping intervals from end to end, so \emph{when} an
order arrives relative to the flow it is following is worth $55\,\%$ of the
passive cost and more than a tripling of the aggressive one. That is only
possible if the placement and cancellation decisions stand in a real temporal
relationship to the order flow --- which is to say that following an identified
order carries information, rather than amounting to posting at a sensible depth
by another route.

The slope measures how tightly a method is coupled to the flow, and the two
styles rank as the mechanism predicts. Aggressive POV must \emph{reach} a price
the flow has just printed, the tightest coupling available, and loses
$0.042$ ticks per contract per millisecond. Passive \pov{} inherits a price
and a lifetime but waits to be reached, and loses $0.011$.

\subsection{A control: the same method on stale information}
\label{sec:stale}

The latency sweep above measures a slope over delays a real participant might
actually face. This subsection is a different kind of experiment and is kept
separate for that reason: thirty seconds is not an operating point, it is a
\emph{control}. The question it answers is how much of \pov{}'s cost advantage
comes from the information it inherits, and it answers it by destroying that
information's freshness while holding everything else fixed.

\begin{table}[!htbp]
\centering
\caption{The same passive configuration run against itself, with and without a
thirty-second delay on the order, the cancel and the feed. ES futures, 245
sessions of calendar 2025 without the delay and 246 with it, 100 contracts
bought and 100 sold per window, the
passive configuration of Table~\ref{tab:styles} throughout. Thirty seconds is
chosen as a control
rather than as a plausible latency: it is long enough that the observation the
shadow was placed from has been overtaken by the market many times over.
Intervals are 95\,\% and clustered by session.}
\label{tab:stale}
\small
\begin{tabular}{lrrrrr}
\toprule
Delay & Windows & Rel.\ slippage & Participation & Time & Child fills \\
 & & (ticks/contract) & (\%) & (s) & (per leg) \\
\midrule
None       & 9{,}554 & $+0.0953 \pm 0.0130$ & 4.47 & 72 & 85.5 \\
30 seconds & 9{,}594 & $+0.3673 \pm 0.0989$ & 1.56 & 44 & 37.2 \\
\bottomrule
\end{tabular}
\end{table}

Cost rises by a factor of $3.9$, on intervals that do not approach each other,
and three further columns move with it: participation falls to a third, the
parent is worked in $37$ child fills per leg rather than $85$, and the legs
finish \emph{sooner} --- because a quote priced thirty seconds ago arrives
marketable and crosses on contact rather than resting, which is item~2 of
Section~\ref{sec:latency} at an extreme. The difference, $0.272$ ticks per
contract, is what acting immediately on an observed order is worth over acting
on the same observation half a minute later.

What this establishes is that
the inherited information is real and that it is \emph{perishable}. It does not
establish what that information is worth against placing with none at all --- at
thirty seconds the price is still inherited from a real participant's order,
merely stale.

The zero-delay row also reproduces the passive column of Table~\ref{tab:styles}
on an independent re-run: $+0.0953$ against $+0.0955$, $4.47\,\%$ participation
against $4.46\,\%$, and 72\,s against 72.7\,s.

\subsection{Threats to validity}

The material limitation is that the replay does not model the market's reaction
to \pov{}'s own orders. The simulator fills the algorithm's
orders against the recorded feed as though those orders had not been there:
a fill is \emph{added} to the day's volume rather than displacing someone else's.
At the $4.5\,\%$ aggregate participation of the zero-latency comparison that
assumption is mild. At the $6.9\,\%$ reached by passive \pov{} at 5\,ms it is strained --- and
individual windows reach far higher --- so the true denominator would be smaller
and the reported participation should be read as an upper bound.

Two further caveats. The study is a single instrument: ES is a one-tick book for
most of the session, and the relative-to-immediate figure in particular is close
to an identity in such a book, so it should not be carried over to wider-spread
products without re-measurement. And the parent size is fixed at 100 contracts
throughout; the size dependence is the subject of the next section.

\section{Participation Rate and Parent-Order Size}
\label{sec:partsize}

\pov{} controls its \emph{order-placement rate} (Section~\ref{sec:select}); it
does not control the participation rate directly, nor the fill of any
one shadow. Because each shadow joins the queue behind the add it
attaches to, while the amount of depth ahead of it is unknown at placement,
whether and when it fills is a random event. A small parent has few fill
opportunities, so its realized participation is highly variable; as the number of
completed child fills grows, averaging reduces that variability. This is a general
property of passive percentage-of-volume and volume-tracking
execution rather than a feature specific to \pov{}: the delivered participation
is a realized quantity, a recognized source of
tracking error and execution risk
\cite{kissell,cjp2015,bialkowski2008,freiwestray2015}. \pov{} inherits this
behavior, with the queue-position uncertainty of the followed order as the
concrete source of per-fill randomness \cite{moallemi2016,negdrift2024}; the
dependence of that dispersion on parent-order size is what the model below
quantifies.

\paragraph{A stylized model.} To isolate the dependence on size, consider
executing a parent of $Q$ one-lot children. Each child attaches to a random add
whose queue position is unknown. For illustration we assume the market volume
that must trade before a child fills is $\mathrm{Exponential}(\rho)$ with mean
$1/\rho$ for a target participation rate $\rho$ (within this stylized model
$\rho$ denotes the participation rate, distinct from the order-placement rate of
Section~\ref{sec:select}). The market volume elapsed when
the parent completes its $Q$ fills is
\[
  M = \sum_{i=1}^{Q} G_i, \qquad G_i \sim \mathrm{Exponential}(\rho)
  \;\;\Longrightarrow\;\; M \sim \mathrm{Gamma}(Q, \rho),
\]
and the realized participation is $\hat{p} = Q/M$. The ratio
$\hat{p}/\rho = Q/\mathrm{Gamma}(Q,1)$ is independent of $\rho$, so the dependence
on $\rho$ cancels and the model predicts the same normalized size dependence for
any $\rho$. The qualitative conclusion --- concentration on the model's mean as
$Q$ grows --- follows from the central limit theorem and holds for a broad class
of finite-variance gap distributions. The independent-exponential gap assumption
is a strong simplification: real fill gaps are correlated across concurrent
shadows, volatility regimes, and clustered aggressive events, and modeling the
children as filling one at a time produces greater participation variance than
allowing multiple concurrent fills to average independently. The model does not
estimate \pov{}'s actual participation --- the exponential fill-gap distribution
and its rate are not fitted to the CME corpus --- so the result is a scaling law
under a stylized model, not an empirical claim that \pov{} achieves a particular
participation rate.

\begin{figure}[t]
\centering
\includegraphics[width=\textwidth]{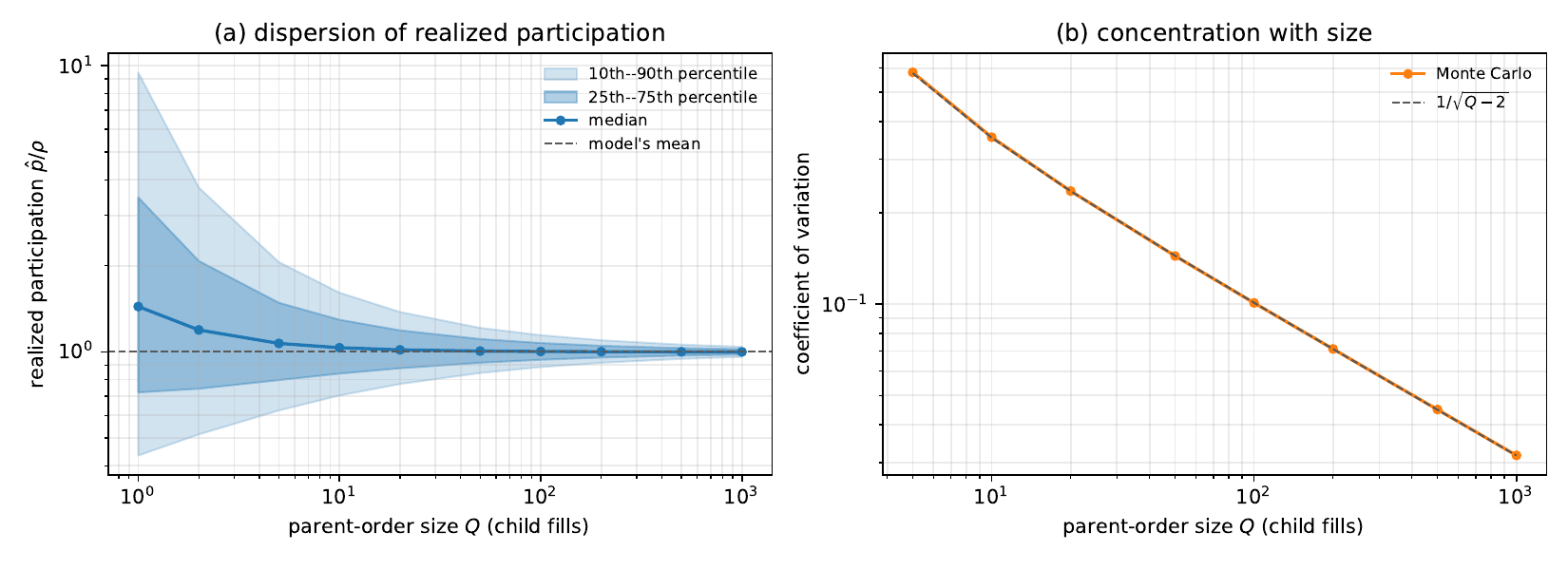}
\caption{Realized participation relative to the model's mean versus parent-order
size $Q$
(number of one-lot child fills), from the stylized model $\hat{p}/\rho =
Q/\mathrm{Gamma}(Q,1)$. (a) Median and percentile bands: the ratio is dispersed
for small $Q$ and concentrates on the model's mean (dashed line) as $Q$ grows.
(b) The coefficient of variation falls as $1/\sqrt{Q-2}$.}
\label{fig:partsize}
\end{figure}

\paragraph{Results.} Figure~\ref{fig:partsize} and Table~\ref{tab:partsize}
report a Monte Carlo of $\hat{p}/\rho$ over $4\times10^{5}$ trials per size. At
$Q=1$ the ratio is $1/\mathrm{Exponential}(1)$: heavy-tailed with a divergent
mean and a 10th--90th percentile band spanning $[0.44, 9.5]$ times the mean --- the
participation is effectively random. As $Q$ grows the ratio concentrates on the
model's mean and its coefficient of variation falls as $1/\sqrt{Q-2}$ (asymptotically $1/\sqrt{Q}$): the 10th--90th
band narrows to $[0.70, 1.61]$ at $Q=10$, $[0.89, 1.14]$ at $Q=100$, and
$[0.96, 1.04]$ at $Q=1000$.

\begin{table}[t]
\centering
\caption{Participation relative to the model's mean versus parent-order size,
$4\times10^{5}$ Monte Carlo trials per size. CV is finite only for $Q>2$.}
\label{tab:partsize}
\small
\begin{tabular}{rrrrr}
\toprule
$Q$ (child fills) & median & 10th pct & 90th pct & CV \\
\midrule
1    & 1.44 & 0.44 & 9.48 & --- \\
2    & 1.19 & 0.51 & 3.77 & --- \\
5    & 1.07 & 0.63 & 2.06 & 0.58 \\
10   & 1.03 & 0.70 & 1.61 & 0.35 \\
20   & 1.02 & 0.77 & 1.38 & 0.24 \\
50   & 1.01 & 0.84 & 1.22 & 0.14 \\
100  & 1.00 & 0.89 & 1.14 & 0.10 \\
200  & 1.00 & 0.92 & 1.10 & 0.07 \\
500  & 1.00 & 0.95 & 1.06 & 0.05 \\
1000 & 1.00 & 0.96 & 1.04 & 0.03 \\
\bottomrule
\end{tabular}
\end{table}

\paragraph{Implication.} The size at which realized participation becomes
concentrated bounds where \pov{}, or any passive follower, delivers a stable participation.
Under the stylized model a parent comprising hundreds or thousands of child fills
has realized participation within a few percent of the model's mean; a very small
order does not, and its participation is
effectively random. For a small parent order the dispersion
can be reduced by modifying \pov{} to shadow only adds reasonably close to the
inside of the book (tightening the placement band of Section~\ref{sec:prod}),
which have a higher and more predictable fill probability; without such a
restriction a small order's participation is dominated by the queue-position
randomness of the few orders it follows, as the small-$Q$ case shows.
This also bounds the proposal of
Section~\ref{sec:disc}: the benchmark role is proposed for parent sizes large
enough that the participation band is narrow, and the width of that band is a
second axis --- alongside slippage --- on which a predictive placement model
that controls queue position could be scored against it.

\section{A benchmark for passive placement}
\label{sec:disc}

\pov{} places passive orders
without an order-book model and without predicting where they will fill, and it
executes at the relative slippage measured in Section~\ref{sec:eval}. We propose
it as a benchmark for passive-order placement on a CLOB rather than as an
optimal method.

\paragraph{Why a model-free passive method should not beat the mid-price.}
The sign of the measured cost is not an accident of this implementation, and it
is worth stating as a prediction rather than reporting as an outcome. In the
Glosten--Milgrom account of the spread \cite{glostenmilgrom1985}, the quoted
width exists to compensate the party supplying liquidity for trading against
possibly better-informed flow: a resting order is filled precisely when somebody
wants to take the other side, so the half-spread it captures is paid back, in
expectation, through adverse selection. A passive method carrying \emph{no}
forecast of where the price is going has nothing with which to separate the
fills it wants from the fills it does not, and should therefore execute slightly
worse than the mid-price --- which is what Section~\ref{sec:eval} measures.

Two consequences follow, and they are what make the method useful as a
reference. First, a passive result at or better than mid from a model-free
method would indicate an error in the measurement or an unmodelled edge, so the
small positive cost is a sanity condition the benchmark must satisfy, not a
weakness to be minimized. Second, the same argument predicts that the expected
\emph{cost} of resting and of crossing should converge, leaving the choice
between them a choice about speed and participation rather than about price. A
predictive placement model is then scored on exactly the quantity \pov{} cannot
address: how much of that adverse-selection cost it recovers by declining the
fills a forecast identifies as toxic. Given the
same passive mandate, a predictive placement model can be scored against
\pov{} to test whether its additional modeling produces a measurable
improvement; the present study reports no such competing baselines, so the
benchmark role is proposed rather than demonstrated.

Adaptive scheduling methods are already evaluated against fixed references: a
reinforcement-learning or optimal-execution scheduler is judged against TWAP,
VWAP, and the Almgren--Chriss trajectory \cite{almgren2000,queuereactive2025},
so any claimed gain is stated relative to a schedule that carries no forecast of
its own. That works because a schedule crossing every slice is fully specified
by its clock. A TWAP that works its slices passively is not: it still has to
decide where in the book each slice rests, and that decision needs the very
placement model the reference was supposed to be free of
(Section~\ref{sec:priorart}). So the fixed reference the literature already has
covers the schedule, and stops at the point the order goes passive.

That is the gap. The naive alternatives at the placement layer --- posting at
the touch, or post-and-wait then cross
\cite{contkukanov2015,limitmarkettactics2014} --- are ad hoc, and each smuggles
in a choice of depth or deadline that the next author will make differently.
\pov{} is well specified in a page, has no free parameter that decides where to
rest, and can be re-run by anyone with the same data. A practitioner proposing a
predictive placement model can then show whether it beats shadow-following, as
an execution scheduler shows whether it beats TWAP.

\paragraph{The bar a placement model has to clear.} Stated plainly, the proposal
is a floor and it has two rungs. \pov{} places passively with no view at all,
and aggressive POV crosses with no view at all; between them they bracket what
is achievable by following flow and nothing else. A designer proposing a
predictive placement model should have to beat \emph{both} on the same corpus,
the same parent, and the same participation: beating \pov{} shows the forecast is
worth something over resting blind, and beating aggressive POV shows it is worth
something over simply taking liquidity. A model that beats neither has bought
complexity and a fitting risk in exchange for nothing, and its author should be
able to see that from two numbers. This is the role we propose --- not a claim
that \pov{} is hard to beat, but a claim that it is the right thing to be
measured against, in the way a scheduler is measured against TWAP.

\pov{} is an execution method: its scope is completing a given position at a
small, predictable cost. When direction is known, an informed aggressive
strategy outperforms it, and \pov{} leaves such directional decisions to the
strategy layer above it.

\section{Future Work}
\label{sec:future}

The measurements of Section~\ref{sec:eval} establish the mechanism and
characterize its cost over one instrument and one year. A completed benchmark
study needs more than that, and several directions remain before \pov{} can be
treated as a validated reference.

\paragraph{Comparison against established methods.} The central open task is to
score \pov{} against competing passive-placement rules under the same
direction-neutral slippage metric: posting at the touch; the passive legs of
TWAP, VWAP, and POV; and at least one \emph{predictive} placement model. Only
such a comparison can establish whether \pov{} is a useful benchmark and by how
much a predictive model improves on it. Section~\ref{sec:stale} establishes that
the inherited information has value and measures how quickly it decays; what a
comparison adds is where the method sits against the alternatives a desk would
otherwise reach for.

\paragraph{Understanding the algorithm's behavior.} Beyond a single cost figure,
the method's behavior needs characterization: realized fill and completion
rates, the fraction of followed orders whose lifecycle ends in a fill rather than
a cancellation, queue-position dynamics, and sensitivity to the order-placement
rate $\rho$ and the grace window $B$.

\paragraph{Ablating the guards.} The safety guards (the two-band hysteresis and the completion and over-hedge
limits) should be ablated to quantify their effect on the reported costs.

\paragraph{Market impact and breadth.} The replay uses a zero-impact simulator;
a controlled experiment that injects synthetic size and measures the perturbation
of subsequent flow would test that assumption. The corpus is a single
index-futures contract; extending to other products, to fragmented multi-venue
equity markets --- where the flow-inherited venue selection of
Section~\ref{sec:sor} would actually be exercised --- and to larger sizes is
required before the results can be generalized.

\section{Conclusion}

We presented \pov{}, a passive execution method that trades the schedules and
aggregate-volume targets of VWAP/TWAP/POV for direct observation of the passive
orders participants actually post. By shadowing individual resting orders by
exchange identifier --- mirroring price and venue, and tying cancellation to
the shadowed order's lifecycle through a lookup on the followed order's identifier 
--- \pov{} works without a volume forecast of its own, uses followed orders as
observable references for local execution flow, and obtains venue selection as
an emergent property of the flow it follows rather than from a routing
computation. Over a full
calendar year of replayed CME ES sessions we compared passive \pov{} against an
aggressive equivalent at nearly matched participation, and swept the simulated latency
from zero to five milliseconds on the order, the cancel and the feed together.

Three results stand out. \emph{The method executes close to the mid.} Over
$9{,}592$ completed passive windows a round trip of 100 contracts bought and 100
sold costs $+0.0955 \pm 0.0130$ ticks per contract against the arrival mid --- under
a tenth of a tick, or $\$1.19$ per contract on ES --- and beats immediate
execution (crossing the spread) by $0.4236$ ticks. 
Post-fill mark-outs computed from different data at
a different grain agree with that figure, so it is a measurement rather than an
artifact of one definition.

\emph{Passive and aggressive cost the same, at zero latency.} At matched
participation the aggressive equivalent returns $+0.0952 \pm 0.0135$ against
passive's $+0.0955$, and $-0.4239$ against $-0.4236$ relative to immediate
execution --- agreement to the fourth decimal on both, over more than nine
thousand windows each. This is what the Glosten--Milgrom account of the spread
predicts of a method carrying no forecast: the half-spread a resting order
captures is returned through adverse selection, and the half-spread a crossing
order pays buys the avoidance of it.

\emph{Latency degrades both, and the method remains usable throughout the range
tested.} Cost rises monotonically with round-trip delay in both styles, with
non-overlapping intervals from end to end. Passive \pov{} goes from $+0.097$ to
$+0.150$ ticks per contract between zero and five milliseconds, still beating
immediate execution by $0.371$ ticks at the far end; aggressive POV more than
triples over the same range, degrading about four times as fast, and the
ordering between the two reverses inside the first half-millisecond.

Taken together these say the two styles are complementary rather than
competitive. They are indistinguishable on price when the algorithm is fast, so
the choice between them is a choice about speed, participation and completion
--- and under delay they fail in opposite directions, passive gaining
participation as its quotes convert into crosses and aggressive losing it as its
marketable orders miss. An implementation that can reach the exchange quickly
has a genuine choice; one that cannot should rest.

Because its placement decision reads a price off an observed order and stops
there, we propose \pov{} as a model-free benchmark for passive-order
placement --- a reference a predictive placement model should be expected to
beat, on the improvement it delivers over these numbers and at a latency it can
actually achieve.

\section*{Reproducibility and disclosure}

Results are produced by a deterministic replay of recorded CME PCAP market data
through open-source C++ actor code (Section~\ref{sec:sim}); a fixed session replays
to identical fills. The \pov{}
implementation and the order-book simulator are available as the
\textsc{kaspar-hft} project.\footnote{\url{https://github.com/vincent212/kaspar-hft}}

\paragraph{Acknowledgments.} The slippage-measurement corpus and the
deterministic replay harness were built at M2 Technologies.


\end{document}